\documentclass{article}%
\usepackage{graphicx}
\usepackage{amssymb}
\usepackage{amsmath}
\usepackage{amsfonts}%
\newtheorem{theorem}{Theorem}

\ifx\pdfoutput\relax\let\pdfoutput=\undefined\fi
\newcount\msipdfoutput
\ifx\pdfoutput\undefined\else
\ifcase\pdfoutput\else
\ifx\paperwidth\undefined\else
\ifdim\paperheight=0pt\relax\else\pdfpageheight\paperheight\fi
\ifdim\paperwidth=0pt\relax\else\pdfpagewidth\paperwidth\fi
\fi\fi\fi
\begin{document}

\title{Spectral parameter power series representation for Hill's discriminant}
\author{K.V. Khmelnytskaya and H.C. Rosu\\{\small IPICYT, Instituto Potosino de Investigacion Cientifica y Tecnologica,}\\{\small Apdo Postal 3-74 Tangamanga, 78231 San Luis Potos\'{\i}, S.L.P., Mexico}\\
{\small khmel@uaq.mx, hcr@ipicyt.edu.mx}}

\date{Annals of Physics 325 (Nov. 2010) pp. 2512-2521\\
{\small arXiv:1002.1514v3}}
\maketitle

\begin{abstract}
{\footnotesize \noindent We establish a series representation of the Hill
discriminant based on the spectral parameter power series (SPPS) recently introduced
by V. Kravchenko. We also show the invariance of the Hill discriminant under
a Darboux transformation and employing the Mathieu case the feasibility of this type of series for
numerical calculations of the eigenspectrum.\\


}

\end{abstract}

\section{Introduction}

There is recent strong interest in differential equations with periodic
coefficients due to their numerous applications in modern material sciences
and engineering. The main goal in their theoretical framework is the
description of the spectrum and getting the periodic and quasiperiodic
solutions. In the case of linear second-order ordinary differential equation,
a function of the spectral parameter known as Hill's discriminant is the basic
quantity containing important information both about the spectrum of the
differential operator and also about the construction of the (quasi)periodic solutions.

This paper focuses on a representation of the Hill discriminant which is
given in the form of a power series in the spectral parameter. This
representation is obtained by using recent results of Kravchenko
\cite{Krav2008,Krav2009,KP}. It is worth mentioning that since the Hill
discriminant for each value of the spectral parameter can be obtained from a
couple of corresponding linearly independent solutions, any available
representation for these solutions leads to a representation of Hill's
discriminant. Nevertheless, in general no easy and practically treatable
representation for Hill's discriminant is known unless for the case of some
well-studied equations such as the Mathieu and the Lam\'{e} equations.

Long ago, Jagerman \cite{Jagerman} introduced and studied in detail the
so-called cardinal series representation of Hill's discriminant. In the famous
book of Magnus and Winkler \cite{Magnus} the Hill discriminant for
Schroedinger type equations is expressed as an infinite determinant involving
the Fourier coefficients of the potential as well as the spectral parameter.
The phase-integral method is used by Fr\"{o}man \cite{Froman} to obtain a
representation involving a matrix whose entries are complicated phase
integrals, while Boumenir \cite{Boum} wrote it in terms of integrals derived
from the inverse spectral theory.

In all the aforementioned series representations the spectral parameter enters
in a quite sophisticated way with all the terms being functions of the
spectral parameter. Instead, our result herein gives the Hill discriminant in
the form of a power series in the spectral parameter with the series
coefficients independent of it and calculated only once, i.e., for a single
value of the spectral parameter. Moreover, its practical implementation is
easy and as an illustration we show in the present work the numerical results
obtained for the Mathieu equation.

We also show that the SPPS representation gives additional information not only
concerning the original equation but also on its Darboux-related partners. As
a corollary, we prove the invariance of the Hill discriminant under the
Darboux transformation under some additional conditions.

\section{Hill's type equations}

The Sturm-Liouville differential equation
\begin{equation}
L\left[  f(x,\lambda)\right]  =-(p(x)f^{\prime}(x,\lambda))^{\prime
}+q(x)f(x,\lambda)=\lambda f(x,\lambda) \label{eq}%
\end{equation}
with $T$-periodic coefficients $p(x)$ and $q(x)$ and real parameter $\lambda$
is known as of Hill type. We first recall some necessary definitions and basic
properties associated with the Eq.~(\ref{eq}) from the Floquet (Bloch)
theory. For more details see, e.g., \cite{Magnus, Eastham}. In what follows we
assume that $p(x)>0$, $p^{\prime}(x)$ and $q(x)$ are continuous bounded functions.

For each $\lambda$ there exists a fundamental system of solutions, i.e., two
linearly independent solutions of (\ref{eq}) $f_{1}(x,\lambda)$ and
$f_{2}(x,\lambda)$ which satisfy the initial conditions%
\begin{equation}
f_{1}(0,\lambda)=1,\quad f_{1}^{\prime}(0,\lambda)=0,\quad f_{2}%
(0,\lambda)=0,\quad f_{2}^{\prime}(0,\lambda)=1. \label{cond}%
\end{equation}
Then the Hill discriminant associated with Eq.~(\ref{eq}) is defined as a
function of $\lambda$ as follows
\[
D(\lambda)=f_{1}(T,\lambda)+f_{2}^{\prime}(T,\lambda).
\]
Employing $D(\lambda)$ one can easily describe the spectrum of the
corresponding equation. Namely, the values of $\lambda$ for which $\left\vert
D(\lambda)\right\vert \leq2$ form the allowed bands or stability intervals
meanwhile the values of $\lambda$ such that $\left\vert D(\lambda)\right\vert
>2$ belong to forbidden bands or instability intervals \cite{Magnus}. The band
edges (values of $\lambda$ such that $\left\vert D(\lambda)\right\vert =2$)
represent the discrete spectrum of the operator, i.e., they are the
eigenvalues of the operator with periodic ($D(\lambda)=2$) or antiperiodic
($D(\lambda)=-2$) boundary conditions.
The eigenvalues $\lambda_{n}$, $n=0,1,2,...$ form an infinite sequence
$\lambda_{0}<\lambda_{1}\leqslant\lambda_{2}<\lambda_{3}...$, and an important
property of the minimal eigenvalue $\lambda_{0}$ is the existence of a
corresponding periodic nodeless solution $f_{0}(x,\lambda_{0})$ \cite{Magnus}.
In general solutions of (\ref{eq}) are not of course periodic, and one of the
important tasks related to Sturm-Liouville equations with periodic
coefficients is the construction of quasiperiodic solutions. In this paper, we
use the matching procedure from \cite{James} for which the main ingredient is
the pair of solutions $f_{1}(x,\lambda)$ and $f_{2}(x,\lambda)$ of (\ref{eq})
satisfying conditions (\ref{cond}). Namely, using $f_{1}(x,\lambda)$ and
$f_{2}(x,\lambda)$ one obtains the quasiperiodic solutions $f_{\pm}%
(x+T)=\beta_{\pm}f_{\pm}(x)$ as follows
\begin{equation}
\label{bloch}f_{\pm}(x,\lambda)=\beta_{\pm}^{n}F_{\pm}(x-nT,\lambda
),\quad\left\{
\begin{array}
[c]{c}%
nT\leq x<(n+1)T\\
n=0,\pm1,\pm2,...~
\end{array}
\right.  .
\end{equation}
where $F_{\pm}(x,\lambda)$ are the so-called self-matching solutions, which
are the following linear combinations\ $F_{\pm}(x,\lambda)=f_{1}%
(x,\lambda)+\alpha_{\pm}f_{2}(x,\lambda)$\ with $\alpha_{\pm}$ being roots of
the algebraic equation $f_{2}(T,\lambda)\alpha^{2}+(f_{1}(T,\lambda
)-f_{2}^{\prime}(T,\lambda))\alpha-f_{1}^{\prime}(T,\lambda)=0$. The Bloch
factors $\beta_{\pm}$ are a measure of the rate of increase (or decrease) in
magnitude of the self-matching solutions $F_{\pm}(x,\lambda)$ when one goes
from the left end of the cell to the right end, i.e., $\beta_{\pm}%
(\lambda)=\frac{F_{\pm}(T,\lambda)}{F_{\pm}(0,\lambda)}$. The values of
$\beta_{\pm}$ are directly related to the Hill discriminant, $\beta_{\pm
}(\lambda)=\frac{1}{2}(D(\lambda)\mp\sqrt{D^{2}(\lambda)-4})$, and obviously
at the band edges $\beta_{+}=\beta_{-}=\pm1$ for $D(\lambda)=\pm2$, correspondingly.

\section{The SPPS series representation of Hill's discriminant}

In this section, we will give an efficient representation for the Hill
discriminant using the method of spectral parameter power series (SPPS)
\cite{Krav2008, KP}, which was used in \cite{pap1,pap2} for studying quantum
mechanical models related to one-dimensional Dirac systems, also for solving
electromagnetic scattering problems \cite{ckko} and for solving inverse
problems \cite{kt}. The SPPS method offers a procedure for constructing the solutions
$f_{1}(x,\lambda)$ and $f_{2}(x,\lambda)$ of (\ref{eq}) which\ satisfy the
initial conditions (\ref{cond}) . This construction is based on the knowledge
of one non-vanishing particular solution $f_{0}(x,\lambda_{0})$ of (\ref{eq})
being bounded on $[0,T]$ together with $\frac{1}{f_{0}(x,\lambda_{0})}$. In
the case of Hill's equation the first eigenvalue $\lambda_{0}$ of the Eq.~(\ref{eq}) generates 
nodeless periodic eigenfunction $f_{0}(x,\lambda_{0})$.
In what follows we\ initially suppose that the value of $\lambda_{0}$ is
known. Note that, it can be obtained by different methods including the same
SPPS method \cite{KP} as we explain in subsection \ref{3.4}.

Given $\lambda_{0}$, we proceed in three steps in order to obtain the
representation of Hill's discriminant:

\begin{quote}

\noindent i) the first one is the construction of a particular nodeless
solution $f_{0}(x,\lambda_{0})$ of (\ref{eq}) which is periodic, i.e.,
$f_{0}(x+T,\lambda_{0})=f_{0}(x,\lambda_{0})$,

\noindent ii) the second one is the construction of the fundamental system of
solutions $f_{1}(x,\lambda)$ and $f_{2}(x,\lambda)$ of (\ref{eq}) for all
values of the parameter $\lambda$,

\noindent iii) the final step is getting the representation of Hill's discriminant.

\end{quote}

We detail each of the steps in the following subsections.


\subsection{The nodeless periodic solution}

\label{3.1}

In order to obtain the nodeless periodic solution $f_{0}(x,\lambda_{0})$ of
(\ref{eq}) for $\lambda=\lambda_{0}$, i.e.,
\begin{equation}
-(p(x)(f_{0}(x))^{\prime})^{\prime}+q(x)f_{0}(x)=\lambda_{0}f_{0}(x)
\label{lambda0}%
\end{equation}
we first construct two linearly independent SPPS solutions of (\ref{lambda0}),
$f_{0,1}(x,\lambda_{0})$ and $f_{0,2}(x,\lambda_{0})$, which satisfy the
initial conditions $f_{0,1}(0,\lambda_{0})=$ $f_{0,2}^{\prime}(0,\lambda
_{0})=1$ and $f_{0,1}^{\prime}(0,\lambda_{0})=f_{0,2}(0,\lambda_{0})=0$
\cite{Krav2008}. These solutions are not necessarily periodic. The procedure
of James \cite{James} allows one to obtain from $f_{0,1}(x,\lambda_{0})$ and
$f_{0,2}(x,\lambda_{0})$ the Floquet type solutions which degenerate to a
single periodic solution $f_{0}(x,\lambda_{0})$ since $\lambda_{0}$ represents
a band edge.

The functions $f_{0,1}(x,\lambda_{0})$ and $f_{0,2}(x,\lambda_{0})$ can be
calculated as follows \cite{Krav2008}%

\begin{equation}
f_{0,1}(x,\lambda_{0})=
{\displaystyle\sum\limits_{even\hspace{0.05in}n=0}^{\infty}}
\,\widetilde{X}_{0}^{(n)}\qquad\text{and}\,\qquad f_{0,2}(x,\lambda_{0})=p(0)
{\displaystyle\sum\limits_{odd\hspace{0.05in}n=1}^{\infty}}
X_{0}^{(n)}, \label{f01 f02}%
\end{equation}
where
\[
\widetilde{X}_{0}^{(0)}\equiv1,\qquad X_{0}^{(0)}\equiv1,
\]
\[
\widetilde{X}_{0}^{(n)}(x)=\left\{
\begin{tabular}
[c]{ll}
$\int_{0}^{x}\widetilde{X}_{0}^{(n-1)}(\xi)(q(\xi)-\lambda_{0})d\xi\qquad$ &
$\text{for an odd }n$\\
& \\
$\int_{0}^{x}\widetilde{X}_{0}^{(n-1)}(\xi)\frac{1}{p(\xi)}d\xi\qquad$ &
$\text{for an even }n$%
\end{tabular}
\right.
\]

\[
X_{0}^{(n)}(x)=\left\{
\begin{tabular}
[c]{ll}%
$\int_{0}^{x}X_{0}^{(n-1)}(\xi)\frac{1}{p(\xi)}d\xi\qquad$ & $\text{for an odd
}n$\\
& \\
$\int_{0}^{x}X_{0}^{(n-1)}(\xi)(q(\xi)-\lambda_{0})d\xi\qquad$ & $\text{for an
even }n$%
\end{tabular}
\ \right.
\]
Now following \cite{James} we obtain the periodic nodeless solution of
(\ref{lambda0})
\begin{align}
f_{0}(x,\lambda_{0})  &  =f_{0,1}(x-nT,\lambda_{0})+\alpha f_{0,2}%
(x-nT,\lambda_{0}),\label{f0}\\
&  \left\{
\begin{array}
[c]{c}%
nT\leq x<(n+1)T\\
n=0,1,2,...~.
\end{array}
\right. \nonumber
\end{align}
where $\alpha=\frac{f_{0,2}^{\prime}(T,\lambda_{0})-f_{0,1}(T,\lambda_{0}%
)}{2f_{0,2}(T,\lambda_{0})}$ \ since $\lambda_{0}$ is the band edge eigenvalue
\cite{James}.

\subsection{Fundamental system of solutions}

\label{3.2}

Once having the function $f_{0}(x,\lambda_{0})$ the solutions $f_{1}%
(x,\lambda)$ and $f_{2}(x,\lambda)$ of (\ref{eq}) and (\ref{cond}) for all values
of the parameter $\lambda$ can be given using the SPPS method once again
\cite{Krav2008}
\begin{align}
f_{1}(x,\lambda)  &  =\frac{f_{0}(x)}{f_{0}(0)}\widetilde{\Sigma}%
_{0}(x,\lambda,\lambda_{0})+p(0)f_{0}^{\prime}(0)f_{0}(x)\Sigma_{1}%
(x,\lambda,\lambda_{0}),\nonumber\\
& \label{f1 f2}\\
f_{2}(x,\lambda)  &  =-p(0)f_{0}(0)f_{0}(x)\Sigma_{1}(x,\lambda,\lambda
_{0}).\nonumber
\end{align}
The summations $\widetilde{\Sigma}_{0}$ and $\Sigma_{1}$ are the spectral
parameter power series
\[
\widetilde{\Sigma}_{0}(x,\lambda,\lambda_{0})=\sum_{\,n=0}^{\infty}
\widetilde{X}^{(2n)}(x)(\Delta\lambda)^{n},\quad\Sigma_{1}(x,\lambda
,\lambda_{0})=\sum_{n=1}^{\infty}X^{(2n-1)}(x)(\Delta\lambda)^{n-1}~,
\]
where $\Delta\lambda=\lambda-\lambda_{0}$ and the coefficients $\widetilde
{X}^{(n)}(x)$, $X^{(n)}(x)$ are given by the following recursive relations
\[
\widetilde{X}^{(0)}\equiv1,\qquad X^{(0)}\equiv1,
\]
%

\begin{equation}
\tilde{X}^{(n)}(x)=
\begin{cases}
\int_{0}^{x}\tilde{X}^{(n-1)}(\xi)f_{0}^{2}(\xi)d\xi\qquad\mathrm{for}%
\,\mathrm{an}\,\mathrm{odd}\,n\\
\\
-\int_{0}^{x}\tilde{X}^{(n-1)}(\xi)\frac{d\xi}{p(\xi)f_{0}^{2}(\xi)}%
\qquad\ \ \ \ \mathrm{for}\,\mathrm{an}\,\mathrm{even}\,n
\end{cases}
\label{K1}
\end{equation}

\bigskip

\begin{equation}
X^{(n)}(x)=
\begin{cases}
-\int_{0}^{x}X^{(n-1)}(\xi)\frac{d\xi}{p(\xi)f_{0}^{2}(\xi)}\qquad
\ \ \ \ \ \mathrm{for}\,\mathrm{an}\,\mathrm{odd}\,n\\
\\
\int_{0}^{x}X^{(n-1)}(\xi)f_{0}^{2}(\xi)d\xi\qquad\ \mathrm{for}
\,\mathrm{an}\,\mathrm{even}\,n~.
\end{cases}
\label{K2}
\end{equation}

One can check by a straightforward calculation that the solutions $f_{1}$ and
$f_{2}$ fulfill the initial conditions (\ref{cond}), for this the following
relations are useful
\begin{equation}
\label{S1}\left(  \widetilde{\Sigma}_{0}(x,\lambda,\lambda_{0})\right)
_{x}^{\prime}=-\frac{\widetilde{\Sigma}_{1}(x,\lambda,\lambda_{0})}%
{p(x)f_{0}^{2}(x)},\ \text{where}\,\,\widetilde{\Sigma}_{1}(x,\lambda
,\lambda_{0})=\sum_{\,n=1}^{\infty}\widetilde{X}^{(2n-1)}(x)(\Delta
\lambda)^{n}%
\end{equation}
and%
\begin{equation}
\label{S2}\left(  \Sigma_{1}(x,\lambda,\lambda_{0})\right)  _{x}^{\prime
}=-\frac{\Sigma_{0}(x,\lambda,\lambda_{0})}{p(x)f_{0}^{2}(x)},\ \text{where}
\,\,\Sigma_{0}(x,\lambda,\lambda_{0})=\sum_{n=0}^{\infty}X^{(2n)}
(x)(\Delta\lambda)^{n}.
\end{equation}
Having obtained the fundamental system of solutions for any value of $\lambda
$, one can apply the construction (\ref{bloch}) in order to obtain the Bloch
solutions which become eigenfunctions for $\lambda$ being eigenvalues.

\subsection{\bigskip Hill's discriminant in SPPS form}

\label{3.3}

Now the Hill discriminant $D(\lambda)=f_{1}(T,\lambda)+f_{2}^{\prime
}(T,\lambda)$ can be written in a simple explicit form. For this we write
$f_{1}(T,\lambda)$ and $f_{2}^{\prime}(T,\lambda)$ in a form of a spectral
parameter power series using (\ref{f1 f2}) and taking into account
(\ref{S2}):
\begin{align}
D(\lambda)  &  =\frac{f_{0}(T)}{f_{0}(0)}\widetilde{\Sigma}_{0}(T,\lambda
,\lambda_{0})+\frac{f_{0}(0)}{f_{0}(T)}\Sigma_{0}(T,\lambda,\lambda
_{0})\label{DD}\\
&  +\left(  f_{0}^{\prime}(0)f_{0}(T)-f_{0}(0)f_{0}^{\prime}(T)\right)
p(0)\Sigma_{1}(T,\lambda,\lambda_{0})~.\nonumber
\end{align}
Finally, taking into account that $f_{0}(x)$ is a $T$-periodic function
$f_{0}(0)=f_{0}(T)$ and\ writing the explicit expressions for $\widetilde
{\Sigma}_{0}(T,\lambda,\lambda_{0})$ and $\Sigma_{0}(T,\lambda,\lambda_{0})$
we obtain a representation for Hill's discriminant associated with (\ref{eq})
\begin{equation}
D(\lambda)\equiv\sum_{n=0}^{\infty}\left(  \tilde{X}^{(2n)}(T)+X^{(2n)}%
(T)\right)  (\Delta\lambda)^{n}\text{.} \label{D}%
\end{equation}
Thus, only one particular nodeless and periodic solution $f_{0}(x,\lambda
_{0})$ of (\ref{eq}) is needed for constructing the associated Hill
discriminant. We formulate the result (\ref{D}) as the following theorem:

\begin{theorem}
Let $\lambda_{0}$ be the lowest eigenvalue of the Sturm-Liouville problem for
the operator $L$ on the segment $[0,T]$ with periodic boundary conditions and
$f_{0}(x,\lambda_{0})$ be the corresponding eigenfunction. Then the Hill
discriminant for (\ref{eq}) has the form (\ref{D}) where $\tilde{X}^{(2n)}$
and $X^{(2n)}$ are calculated according to (\ref{K1}) and (\ref{K2}), and the
series converges uniformly on any compact set of values of $\lambda$.
\end{theorem}

To illustrate the formula (\ref{D}) we consider a simple example. Let in
equation (\ref{eq}) $q(x)=0$, $p(x)=1$. It is easy to see that the associated
discriminant is $D(\lambda)=2\cos\sqrt{\lambda}T$, from where we obtain
$\lambda_{0}=0$\ and a corresponding non-trivial periodic solution is
$f_{0}(x)=1$. Now making use of this solution we construct the discriminant by
means of the formula (\ref{D}). The coefficients $\tilde{X}^{(2n)}(T)$ and
$X^{(2n)}(T)$ given by (\ref{K1}) and (\ref{K2}) take the form
\[
\tilde{X}^{(2n)}(T)=X^{(2n)}(T)=(-1)^{n}\frac{T^{2n}}{(2n)!},\quad
n=0,1,2,...~.
\]
The substitution in (\ref{D}) gives $D(\lambda)=2\cos\sqrt{\lambda}T$.

\subsection{Construction of the first eigenvalue $\lambda_{0}$ by the SPPS method}

\label{3.4}

Notice that in the expression (\ref{DD}) for $D(\lambda)$ and in all
reasonings previous to it we do not use the periodicity of the solution
$f_{0}(x,\lambda_{0})$, therefore (\ref{DD}) and the whole procedure for
obtaining it are valid for any $\lambda_{\ast}$ such that there exists a
corresponding solution $f_{\ast}(x,\lambda_{\ast})$ which is bounded on
$[0,T]$\ together with $1/(pf_{\ast}^{2})$. Such a solution $f_{\ast
}(x,\lambda_{\ast})$ can be obtained in the following way
\begin{equation}
f_{\ast}(x,\lambda_{\ast})=f_{\ast,1}(x,\lambda_{\ast})+if_{\ast,2}%
(x,\lambda_{\ast}) \label{f_asterisc}%
\end{equation}
where $f_{\ast,1}(x,\lambda_{\ast})$ and $f_{\ast,2}(x,\lambda_{\ast})$ are
given by (\ref{f01 f02}) with $\lambda_{\ast}$ instead of $\lambda_{0}$. For
more details see \cite{Krav2009}. The pair of the independent solutions
$f_{1}(x,\lambda)$ and $f_{2}(x,\lambda)$ of (\ref{eq}) given by (\ref{f1 f2})
of course are independent of the choice of the solution $f_{0}(x,\lambda_{0}%
)$, hence instead of $f_{0}(x,\lambda_{0})$ in (\ref{DD}) one can take
$f_{\ast}(x,\lambda_{\ast})$ given by (\ref{f_asterisc}). \ Thus, in terms of
$f_{\ast}(x,\lambda_{\ast})$ where $\lambda_{\ast}$ is essentially arbitrary,
$D(\lambda)$ can be represented as a series in powers of $(\lambda
-\lambda_{\ast})$%
\begin{align}
D(\lambda)  &  =\sum_{n=0}^{\infty}\left(  \frac{f_{\ast}(T)}{f_{\ast}%
(0)}\tilde{X}^{(2n)}(T)+\frac{f_{\ast}(0)}{f_{\ast}(T)}X^{(2n)}(T)+\right.
\label{Dstar}\\
&  +\left(  f_{\ast}^{\prime}(0)f_{\ast}(T)-f_{\ast}(0)f_{\ast}^{\prime
}(T)\right)  p(0)X^{(2n+1)}(T)%
\genfrac{.}{)}{0pt}{}{{}}{{}}
(\lambda-\lambda_{\ast})^{n}\text{.}\nonumber
\end{align}

Now the band edge $\lambda_{0}$ required for the formula (\ref{D}) can be
calculated as a first zero of the expression $D(\lambda)-2$ where $D(\lambda)$
is given by (\ref{Dstar}). For the numerical purpose it can be useful to know
the interval containing $\lambda_{0}$. Since $q$ is a bounded periodic function,
there is a number $\Lambda$ which satisfies the inequality $q(x)>\Lambda$
$\forall x\in\mathbf{R}$. It is known \cite{Eastham} that $D(\lambda)>2$ for
all $\lambda\leqslant\Lambda$, therefore the lower estimate for $\lambda_{0}$
is the following
\[
\lambda_{0}\geqslant\min q(x).
\]
The upper bound can be obtained considering the Rayleigh quotient for periodic
problem \cite{Pinchover}
\[
\lambda_{0}\leqslant\frac{\int_{0}^{T}\left(  p\left(  x\right)  (u^{\prime
}\left(  x\right)  )^{2}+q\left(  x\right)  \left(  u(x)\right)  ^{2}\right)
dx}{\int_{0}^{T}\left(  u(x)\right)  ^{2}dx}~,
\]
where $u(x)\in\mathbf{C}^{2}\left[0,T\right]$ is periodic with period $T$. The equality occurs if and
only if $u(x)$ is an eigenfunction corresponding to $\lambda_{0}$.

\section{Hill's discriminant of the supersymmetric (SUSY)-related equation}

In this section, we consider the SUSY-related equation of Eq.~(\ref{eq})
and obtain its SPPS solutions. These solutions allow us to prove the equality
between the Hill discriminants of equation (\ref{eq}) and its SUSY-related
Eq.~(\ref{eq1}). For various aspects of SUSY periodic problems, see
\cite{Samsonov,Correa,Cooper}.

The left-hand side of the equation (\ref{eq}) can be factorized in the
following way \cite{Plastino}
\begin{equation}
L\left[  f(x,\lambda)\right]  =\left(  -d_{x}p^{\frac{1}{2}}(x)+\Phi
(x)\right)  \left(  p^{\frac{1}{2}}(x)d_{x}+\Phi(x)\right)  f(x), \label{fact}%
\end{equation}
where $d_{x}$ means the $x$-derivative, the superpotential $\Phi(x)$ is
defined as follows $\Phi(x)=-p^{\frac{1}{2}}(x)\frac{f_{0}^{\prime}%
(x,\lambda_{0})}{f_{0}(x,\lambda_{0})}$. Using this factorization the
coefficient $q(x)$ can be expressed as
\[
q(x)=\Phi^{2}(x)-\left(  p^{\frac{1}{2}}(x)\Phi(x)\right)  ^{\prime}%
+\lambda_{0}.
\]
Introducing the following Darboux transformation
\begin{equation}
\left(  p^{\frac{1}{2}}(x)d_{x}+\Phi(x)\right)  f(x,\lambda)=\tilde
{f}(x,\lambda), \label{Darb}%
\end{equation}
one obtains the equation supersymmetrically related to equation (\ref{eq})
\[
\tilde{L}\left[  \tilde{f}(x,\lambda)\right]  =\left(  p^{\frac{1}{2}}%
(x)d_{x}+\Phi(x)\right)  \left(  -d_{x}p^{\frac{1}{2}}(x)+\Phi(x)\right)
\tilde{f}(x,\lambda)=\lambda\tilde{f}(x,\lambda),
\]
which can be written as follows
\begin{equation}
-d_{x}(p(x)d_{x}\tilde{f}(x,\lambda))+\tilde{q}(x)\tilde{f}(x,\lambda
)=\lambda\tilde{f}(x,\lambda), \label{eq1}%
\end{equation}
where $\tilde{q}(x)$ is the SUSY partner of the potential $q(x)$ given by%

\begin{equation}
\tilde{q}(x)=q(x)+2p^{\frac{1}{2}}(x)\Phi^{\prime}(x)-p^{\frac{1}{2}%
}(x)(p^{\frac{1}{2}}(x))^{\prime\prime}. \label{q2}
\end{equation}
It is worth noting that as $\Phi(x)$ is a $T$-periodic function, the Darboux
transformation assures the $T$-periodicity of $\tilde{q}(x)$.

The pair of linearly independent solutions $\tilde{f}_{1}(x,\lambda)$ and
$\tilde{f}_{2}(x,\lambda)$\ of (\ref{eq1}) can be obtained directly from the
solutions (\ref{f1 f2}) by means of the Darboux transformation (\ref{Darb}).
We additionally take the linear combinations in order that the solutions
$\tilde{f}_{1}(x,\lambda)$ and $\tilde{f}_{2}(x,\lambda)$\ satisfy the initial
conditions $\tilde{f}_{1}(0,\lambda)=\tilde{f}_{2}^{\prime}(0,\lambda)=1$ and
$\tilde{f}_{1}^{\prime}(0,\lambda)=\tilde{f}_{2}(0,\lambda)=0$
\begin{align}
\tilde{f}_{1}(x,\lambda)  &  =\frac{p^{\frac{1}{2}}(0)f_{0}(0)}{p^{\frac{1}%
{2}}(x)f_{0}(x)}\Sigma_{0}(x,\lambda)+\frac{[p^{\frac{1}{2}}(x)]^{\prime
}|_{x=0}-\Phi(0)}{\left(  \Delta\lambda\right)  f_{0}(0)p^{\frac{1}{2}%
}(x)f_{0}(x)} \widetilde{\Sigma}_{1}(x,\lambda),\label{g1}\\
\tilde{f}_{2}(x,\lambda)  &  =\frac{p^{\frac{1}{2}}(0)}{\left(  \Delta
\lambda\right)  f_{0}(0)p^{\frac{1}{2}}(x)f_{0}(x)}\widetilde{\Sigma}%
_{1}(x,\lambda). \label{g2}%
\end{align}

These two solutions allow us to write the expression for Hill's discriminant
associated to the equation (\ref{eq1}), that is $\widetilde{D}(\lambda
)=\tilde{f}_{1}(T,\lambda)+\tilde{f}_{2}^{\prime}(T,\lambda)$. We consider
first the derivative of $\tilde{f}_{2}(x,\lambda)$ and evaluate it for $x=T$
\[
\tilde{f}_{2}^{\prime}(x,\lambda)|_{x=T}=-p^{\frac{1}{2}}(0)\frac{[p^{\frac
{1}{2}}(x)]^{\prime}|_{x=T}f_{0}(T)+ p^{\frac{1}{2}}(T)f_{0}^{\prime}%
(T)}{\left(  \Delta\lambda\right)  f_{0}(0)p(T)f_{0}^{2}(T)}\widetilde{\Sigma
}_{1}(T,\lambda)+\frac{p^{\frac{1}{2}}(0)f_{0}(T)}{p^{\frac{1}{2}}(T)f_{0}%
(0)}\widetilde{\Sigma}_{0}(T,\lambda).
\]
Notice that as the functions $f_{0}(x,\lambda_{0})$ and $p(x)$ are
$T$-periodic, i.e., $f_{0}(0,\lambda_{0})=f_{0}(T,\lambda_{0})$ and
$p(0)=p(T)$, then obviously, the functions $f_{0}^{\prime}(x,\lambda
_{0}),p^{\frac{1}{2}}(x)$ and $[p^{\frac{1}{2}}(x)]^{\prime}$ possess the same
properties. Therefore we have
\begin{align*}
\widetilde{D}(\lambda)  &  =\Sigma_{0}(T,\lambda)+\widetilde{\Sigma}%
_{0}(T,\lambda)+\left(  \frac{[p^{\frac{1}{2}}(x)]^{\prime}|_{x=0}-\Phi
(0)}{\left(  \Delta\lambda\right)  f_{0}(0,\lambda_{0})p^{\frac{1}{2}}%
(T)f_{0}(T,\lambda_{0})}-\right. \\
&  \left.  -\frac{[p^{\frac{1}{2}}(x)]^{\prime}|_{x=T}f_{0}(T,\lambda
_{0})+p^{\frac{1}{2}}(T)f_{0}^{\prime}(T,\lambda_{0})}{\left(  \Delta
\lambda\right)  f_{0}(0,\lambda_{0})p^{\frac{1}{2}}(T)f_{0}^{2}(T,\lambda
_{0})}\right)  \widetilde{\Sigma}_{1}(T,\lambda).
\end{align*}
The substitution $\Phi(0)=-p^{\frac{1}{2}}(0)\frac{f_{0}^{\prime}%
(0,\lambda_{0})}{f_{0}(0,\lambda_{0})}$ clearly shows that the expression in
brackets vanishes. Therefore, we obtain
\[
\widetilde{D}(\lambda)=\Sigma_{0}(T,\lambda)+\widetilde{\Sigma}_{0}%
(T,\lambda)=\sum_{n=0}^{\infty}\left(  \tilde{X}^{(2n)}(T)+X^{(2n)}(T)\right)
(\Delta\lambda)^{n}.
\]
Comparing with (\ref{D}) we have the identity
\begin{equation}
D(\lambda)\equiv\widetilde{D}(\lambda). \label{DD1}
\end{equation}

Thus, we have proven the following statement:


\begin{theorem}
Let $\lambda_{0}$ be the first eigenvalue of (\ref{eq})and $f_{0}(x,\lambda_{0})$ the corresponding $T$-periodic 
nodeless eigenfunction. Then the Darboux transformation (\ref{Darb}) with $\Phi(x)=-p^{\frac{1}{2}}%
(x)\frac{f_{0}^{\prime}(x,\lambda_{0})}{f_{0}(x,\lambda_{0})}$ leads to a SUSY-related Eq.~(\ref{eq1}) with the preservation of the Hill discriminant,
i.e., Eq.~(\ref{DD1}) holds.
\end{theorem}

From the identity of discriminants (\ref{DD1}) it is clear that $\lambda_{0} $
gives rise to a nodeless periodic solution $\tilde{f}_{0}(x,\lambda_{0})$ of
Eq.~(\ref{eq1}). Taking $\lambda=\lambda_{0}$ in (\ref{g1}) and
(\ref{g2}) we get this eigenfunction in the form $\tilde{f}_{0}(x,\lambda
_{0})=\frac{1}{p^{\frac{1}{2}}(x)f_{0}(x,\lambda_{0})}~.$

Notice that, the factorization method can be applied to Eq.~(\ref{eq1})
with the superpotential $\Phi_{1}(x)=-p^{\frac{1}{2}}(x)\frac{\tilde{f}_{0}^{^{\prime}}(x,\lambda_{0})}{\tilde{f}_{0}(x,\lambda_{0})}$. 
In this case, we obtain the representation
\[
\tilde{q}=\Phi_{1}^{2}(x)-\left(  p^{\frac{1}{2}}(x)\Phi_{1}(x)\right)
^{\prime}+\lambda_{0},
\]
which reduces to the equality (\ref{q2}) if one notices the relationship
$\Phi_{1}(x)=(p^{\frac{1}{2}}(x))^{\prime}-\Phi(x)$. It can be also shown that
$\tilde{\tilde q}\equiv q$, where $\tilde{\tilde q}=\tilde{q}(x)+2p^{\frac
{1}{2}}(x)\Phi_{1}^{\prime}(x)-p^{\frac{1}{2}}(x)(p^{\frac{1}{2}}%
(x))^{\prime\prime}$ is the superpartner potential of $\tilde{q}(x)$. Thus, the
Darboux transformation (\ref{Darb}) with the superpotential $\Phi_{1}(x)$
applied to Eq.~(\ref{eq1}) does not produce a different potential.

\section{Numerical calculation of eigenvalues based on the SPPS form of Hill's
discriminant}

As is well known, see e.g., \cite{Magnus}, the zeros of the functions
$D(\lambda)\pm2$ represent \ eigenvalues of the corresponding operator. In
this section, we show that besides other possible applications the
representation (\ref{D}) gives us an efficient tool for the calculation of the
discrete spectrum of a periodic Sturm-Liouville operator.

The first step of the numerical realization of the method consists in
calculation of the minimal eigenvalue $\lambda_{0}$ by means of the procedure
given in subsection \ref{3.4} and subsequently in construction of the
corresponding nodeless periodic solution $f_{0}(x,\lambda_{0})$ using formula
(\ref{f0}). The next step of the algorithm is to compute the functions
$\tilde{X}^{(n)}$ and $X^{(n)}$ given by (\ref{K1}) and (\ref{K2}),
respectively. This construction is based on the eigenfunction $f_{0}%
(x,\lambda_{0})$. Finally, by truncating the infinite series for $D(\lambda
)$\ (\ref{D}) we obtain a polynomial in $\Delta\lambda$
\begin{align}
D_{N}(\lambda) &  =\sum_{n=0}^{N}\left(  \tilde{X}^{(2n)}(T)+X^{(2n)}%
(T)\right)  (\Delta\lambda)^{n}\label{DN}\\
&  =2+\sum_{n=1}^{N}\left(  \tilde{X}^{(2n)}(T)+X^{(2n)}(T)\right)
(\Delta\lambda)^{n}.\nonumber
\end{align}
The roots of the polynomials $D_{N}(\lambda)\pm2$ give us eigenvalues
corresponding to Eq.~(\ref{eq}) with periodic and antiperiodic boundary conditions.

As an example, we consider the Mathieu equation with the following
coefficients
\[
p(x)=1,\quad q(x)=2r\cos2x.
\]
The algorithm was implemented in Matlab 2006. The recursive integration
required for the construction of $\tilde{X}_{0}^{(n)}$, $X_{0}^{(n)}$,
$\tilde{X}^{(n)}$ and $X^{(n)}$ was done by representing the integrand through
a cubic spline using the $\emph{spapi}$ routine with a division of the
interval $[0,T]$ into $7000$ subintervals and integrating using the
$\emph{fnint}$ routine. Next, the zeros of $D_{N}(\lambda)\pm2$\ were
calculated by means of the $\emph{fnzeros}$ routine.

In the following tables, the Mathieu eigenvalues were calculated employing the
SPPS \ representation (\ref{D}) for two values of the parameter $r$.
For comparison the same eigenvalues from the National Bureau of Standards (NBS) tables are also displayed
\cite{NBS}.

\medskip

$%
\begin{tabular}
[b]{|l|l|l|}\hline
& $r=1$ & $r=1$ \\\hline
$n$ & $\lambda_{n}\ \text{(SPPS\thinspace)}$ & $\lambda_{n}%
\ \text{(NBS\thinspace)}\,$\\\hline
$0$ & $-0.455139055973837$ & $-0.45513860$\\\hline
$1$ & $-0.110248420387377$ & $-0.11024882$\\\hline
$2$ & $1.859107160521687$ & $1.85910807$\\\hline
$3$ & $3.917024962694820$ & $3.91702477$\\\hline
$4$ & $4.371299312651704$ & $4.37130098$\\\hline
$5$ & $9.047736927007582$ & $9.04773926$\\\hline
$6$ & $9.078369587941564$ & $9.07836885$\\\hline
$7$ & $16.033018848985410$ & $16.03297008$\\\hline
$8$ & $16.033785039658117$ & $16.03383234$\\\hline
$9$ & $25.020598536509114$ & $25.02084082$\\\hline
$10$ & $25.021087773318282$ & $25.02085434$\\\hline
\end{tabular}
\qquad$

$
\begin{tabular}
[b]{|l|l|l|}\hline
& $r=5$ & $r=5$ \\\hline
$n$ & $\lambda_{n}\ \text{(SPPS\thinspace)}$ & $\lambda_{n}
\ \text{(NBS\thinspace)}\,$\\\hline
$0$ & $-5.800045777242780$ & $-5.80004602$\\\hline
$1$ & $-5.790080596840196$ & $-5.79008060$\\\hline
$2$ & $1.858191484309548$ & $1.85818754$\\\hline
$3$ & $2.099460384254221$ & $2.09946045$\\\hline
$4$ & $7.449142541577460$ & $7.44910974$\\\hline
$5$ & $9.236327731534002$ & \\\hline
$6$ & $11.548906947651728$ & \\\hline
$7$ & $16.648219815375526$ & \\\hline
$8$ & $17.096668282587867$ & \\\hline
$9$ & $25.510753265631860$ & $25.51081605$\\\hline
$10$ & $25.551677357240167$ & $25.54997175$\\\hline
\end{tabular}
$

Figures 1 and 2 display the plots of the calculated Hill discriminants for two
values of the Mathieu parameter.%

\begin{figure}[x]
 \centering
 \includegraphics[width= 16.5 cm, height=6.5 cm]{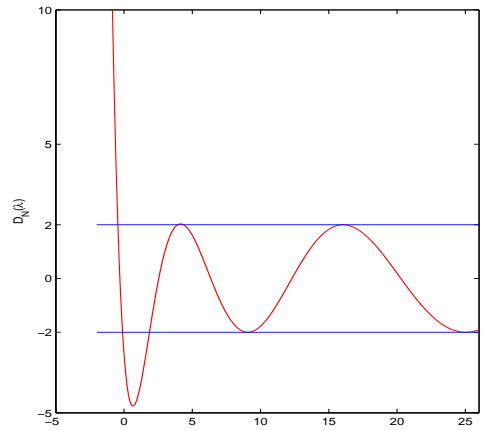}
\caption{\sl The polynomial $D_{N}(\lambda)$ for the Mathieu equation with the
parameter $r=1$ calculated by means of formula (\ref{DN}) for  $N=100$.}
 \label{mathi1}
 \end{figure}

\begin{figure}[x]
 \centering
 \includegraphics[width= 16.5 cm, height=6.5 cm]{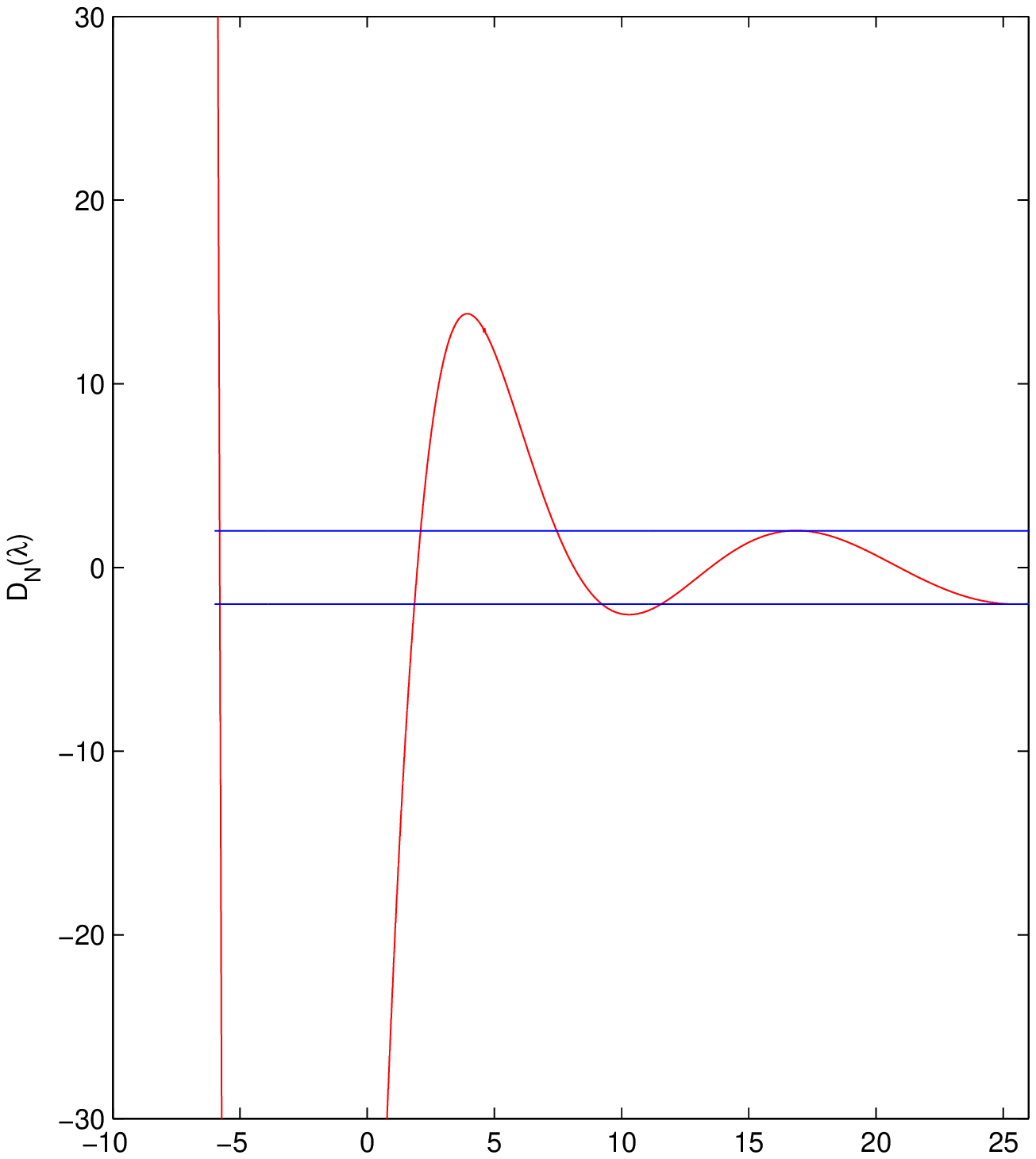}
\caption{\sl  Same as in the previous figure but for $r=5$. The first minimum goes down to -292.0066.}
 \label{mathi2}
 \end{figure}

\section{Conclusions}

In summary, in the present work we obtained a representation of the Hill
discriminant in the form of a series in powers of the spectral distance with
respect to the first eigenvalue
and also tested its efficiency for numerical calculations of the spectral
bands using the Mathieu case.
Moreover, we demonstrate that the Hill discriminant is invariant under the
Darboux transformation generated by the first eigenfunction.

\bigskip

\noindent {\bf Acknowledgment}. The first author thanks CONACyT for a postdoctoral fellowship
allowing her to work in IPICyT.

\end{document}